\documentclass{article}
\usepackage{spconf,amsmath,graphicx}
\usepackage{multirow} 
\usepackage{hyperref} 
\usepackage{multirow} 
\usepackage{amsmath} 
\usepackage{amssymb} 
\usepackage{tabularx,booktabs,caption}
\usepackage{nicematrix}
\newcolumntype{Z}{>{\raggedright}X}

\usepackage[normal]{subfigure}
\usepackage{graphicx}
\usepackage{amsmath}

\renewcommand{\footnotesize}{\scriptsize}
\usepackage{makecell}

\title{Leveraging Speaker Embeddings with Adversarial Multi-task Learning for Age Group Classification}

\name{Kwangje Baeg$^{\star}$ \qquad Yeong-Gwan Kim$^{\dagger}$ \qquad Young-Sub Han$^{\star}$ \qquad Byoung-Ki Jeon$^{\star}$}

\address{$^{\star}$ LG Uplus Corp. \qquad  $^{\dagger}$ LG AI Research}

\begin{document}
\ninept

\maketitle
\begin{abstract}

    Recently, researchers have utilized neural network-based speaker embedding techniques in speaker-recognition tasks to identify speakers accurately. However, speaker-discriminative embeddings do not always represent speech features such as age group well. In an embedding model that has been highly trained to capture speaker traits, the task of age group classification is closer to speech information leakage. Hence, to improve age group classification performance, we consider the use of speaker-discriminative embeddings derived from adversarial multi-task learning to align features and reduce the domain discrepancy in age subgroups. In addition, we investigated different types of speaker embeddings to learn and generalize the domain-invariant representations for age groups. Experimental results on the \textit{VoxCeleb Enrichment} dataset verify the effectiveness of our proposed adaptive adversarial network in multi-objective scenarios and leveraging speaker embeddings for the domain adaptation task.

\end{abstract}
\begin{keywords}
    Adaptive adversarial network, multi-task learning, speaker-discriminative embeddings, age group classification
\end{keywords}

\section{INTRODUCTION}
\label{sec:introduction}

The task of speaker recognition (SR) is the task of identifying or confirming the identity of a person given speech segments \cite{xiang2019margin}. Recently, speaker embeddings learned by deep neural network (DNN)-based architectures such as x-vector and ResNet have shown more impressive performance on SR than the previous state-of-the-art method, the i-vector-based approach \cite{wang2019knowledge}. The DNN-based approach can extract speaker-discriminative and robust speaker embeddings by training on various utterances from a large-scale SR dataset \cite{sang2021deaan}. However, DNN-based speaker-discriminative embeddings do not represent the speech feature itself and cannot be effectively used as the feature vector to analyze speech. Many meaningful details about the speaker's identity such as age, gender, and emotional state are contained in the speech segment \cite{kalluri2020automatic}. However, in practice, neural speaker-discriminative embeddings cannot incorporate speech features into an embedding, except in the SR task, and may simply interpolate the source dataset or even make it impossible to derive a suitable mapping for other speech tasks.

Current speaker-discriminative embeddings are normally trained in the process of aggregating the frame-level features and projected into higher-order nonlinear spaces \cite{sankala2022multi}. The speaker embeddings are well-designed to train the distributed features throughout the utterance and are modeled to extract the rich information as a fixed-length output. However, there is a limit to how robust the features can be made and how well complementary features can be used to improve generalization in downstream tasks. To alleviate these problems, many efforts have been made to improve the representations of the speech segments, facilitate information sharing among different representations, and capture more task-specific evidence from them \cite{sankala2022multi, li2019speaker, zheng2007integration}.

One approach used in \cite{li2019speaker} improves the applicability of speaker-discriminative embeddings by extracting two acoustic features to learn more complementary representations from different acoustic features. Another approach used in \cite{sankala2022multi} enriches the speaker information by incorporating and reconstructing the extracted speaker embeddings while eliminating irrelevant information. Despite such advances in representation techniques, there has been limited research conducted on sharing the information of speaker embeddings with different downstream tasks, such as age group classification. In addition, the task of estimating the precise age and determining the adjacent age group remains a huge challenge by itself, i.e., speech from speakers in adjacent age groups are indistinguishable \cite{fullgrabe2015age, si2022towards}. In this paper, we consider whether speaker-discriminative embeddings can be used for the challenging task of age group classification by leveraging source speaker embeddings.
\begin{figure}[t!]
    \centering
    \includegraphics[width=1.02\columnwidth]{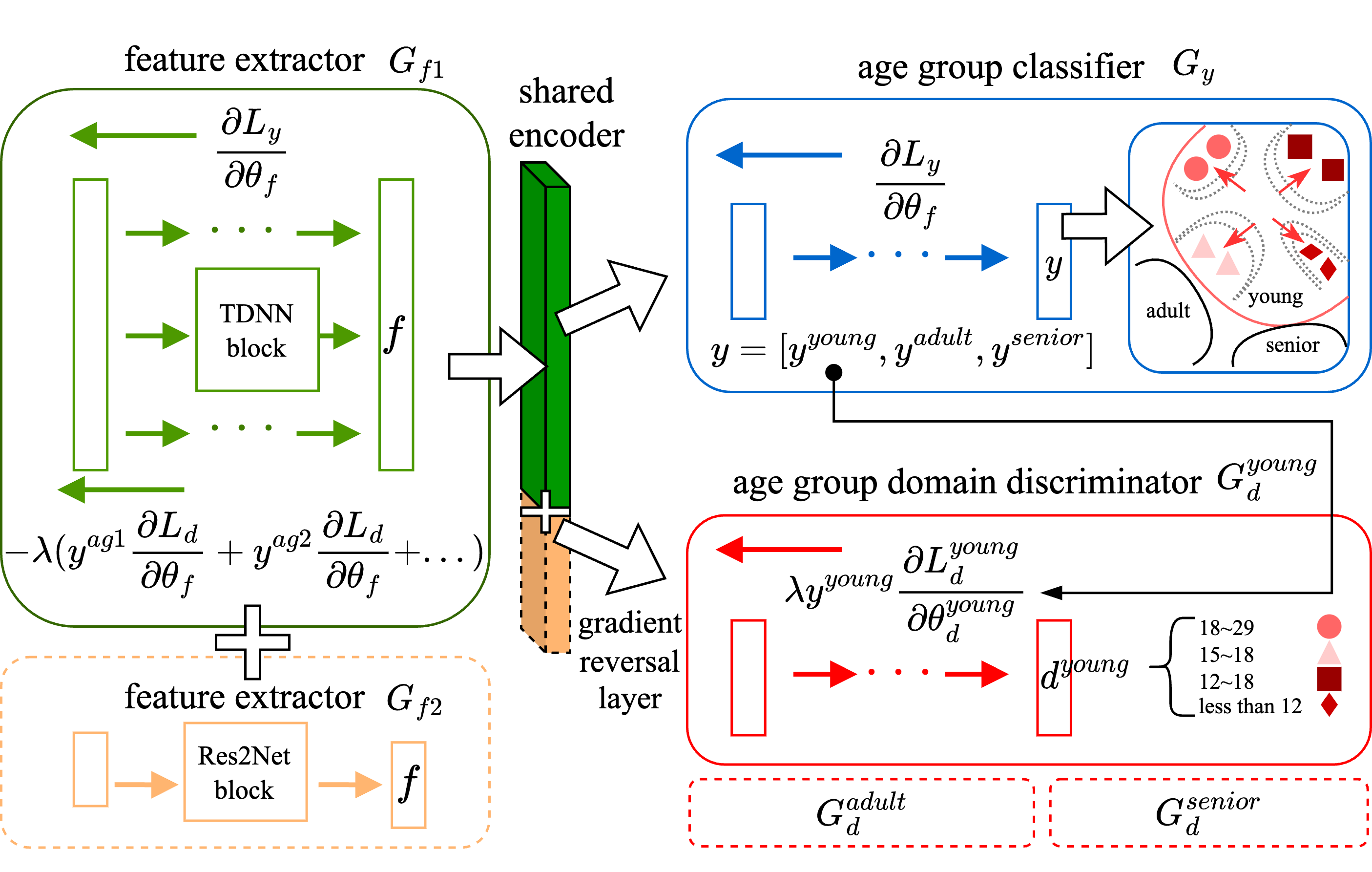}
    \vspace{-1mm} 
    \caption{
        The two main ideas of increasing the suitability of speaker-discriminative embeddings: (a) adaptive adversarial network, (b) integration of embeddings. This example shows for aligning the distributions between age subgroups in \textit{young}-label.
    }
    \label{fig:Architecture}
\end{figure}

We focus on adversarial multi-task representation learning methods \cite{mao2020adaptive, liu2018multi, chen2018gradnorm} to learn the domain-invariant features that represent the age group in speaker embeddings by adversarially training the feature extractor and the domain discriminator. Domain adaptation (DA) aims to reduce the domain difference between the source and target domains to adaptively transfer useful knowledge across domains \cite{li2021feature}. Moreover, the source and target domains are projected onto the same space such that they have different probability distributions. DA is conventionally used to align a rich labeled source domain in a beneficial way for an unlabeled target domain \cite{ganin2015unsupervised}. However, the features extracted from the same speech segment can also be applied to each domain, even if they have different distributions, and these features collectively encode both speaker-related and speaker-unrelated information. An adaptive adversarial network can represent the domain-invariant information with different distributions that transfers well from the original tasks to other tasks \cite{yosinski2014transferable}. To this end, adversarial multi-task learning can enhance the learning of transferable features that reduce the distribution discrepancy between speaker-discriminative features and domain-invariant features \cite{mao2020adaptive, xia2021adaptive}. In our work, a traditional DNN-based age group classifier is adapted by adding a new domain discriminator branch and then trained using the standard domain adversarial training strategy.

In this paper, we propose an adaptive adversarial network architecture that maps well-trained source speaker-discriminative embeddings into a target domain for age group classification. We applied it to a multi-task learning architecture to improve age group classification by leveraging speaker-discriminative embeddings. Experimental results show that the proposed architecture is effective at leveraging speaker embeddings for downstream speech tasks. Moreover, the combination of the feature extractor and discriminator can vary depending on the characteristics of the embeddings. We consider how the adaptive adversarial network can be operated more effectively on speaker-discriminative embeddings without inferring domain-invariant features.

\section{Adaptive Adversarial Networks and Integration of speaker embeddings}
\label{sec:ARCHITECTURE}

\subsection{Adaptive adversarial networks}

\begin{figure}[t!]
    \centering
    \includegraphics[width=1.0\columnwidth]{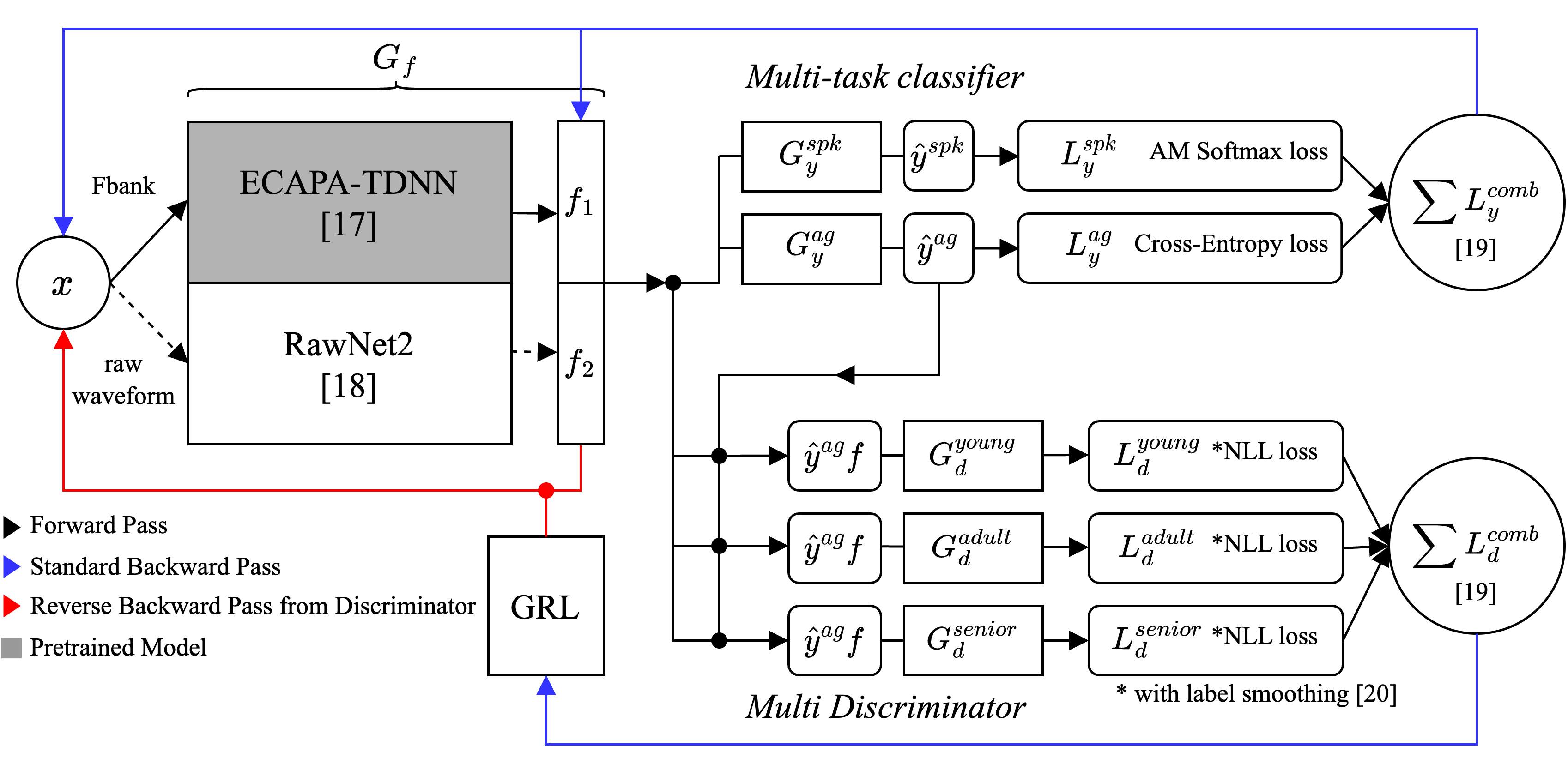}
    \vspace{-3mm} 
    \caption{
        The architecture of the proposed adaptive adversarial network using the integration of speaker-discriminative embeddings, where $G_f$ is the integration of feature extractors, $G^{spk}_y$ and $G^{age}_y$ are label predictors for multi-task learning, and $G^{young}_d$,  $G^{adult}_d$ and $G^{senior}_d$ are domain discriminators for adversarial learning; GRL stands for gradient reversal layer.
    }
    \label{fig:arch}
\end{figure}

The overall model architecture is presented in Fig \ref{fig:arch}. Our adaptive adversarial network is composed of three components: feature extractor $G_f$, label predictor $G_y$, and domain discriminators $G_d$. Feature extractor $G_f$ attempts to generate a domain-invariant feature $f$ to confuse the $G_d$, whereas the discriminators attempt to distinguish the source from the target. The parameters of the feature extractor, label predictors, domain discriminators are denoted by $\theta_f$, $\theta_{y}$, and $\theta_{d}$, respectively. Each data point has three labels: speaker-label and age group-label $x_i \in {D_s}$, and age subgroup-label $x_i \in {D_t}$. $\hat{y}^k_i$ is the softmax output of the age group label predictor $G^{ag}_y$ for each data point $x_i$, which is a probability indicating the degree to which each data point $x_i$ should be attended to the $k$-th domain discriminator $G^k_d$. The objective of adaptive adversarial network can be formulated as:

\begin{equation}
    \begin{aligned}
        \mathcal{L} (\theta_{f}, \theta_{y}, {\theta^{k}_{d}}|^{K}_{k=1} )
        =
        \frac{1}{n_s} \sum_{ {x_i}\in{D_s} } \mathcal{L}_y(G_y (G_f(x_i; \theta_{f}); \theta_{y}), y_i) \\
        - \frac{\lambda}{n_t} {\sum_{k=1}^{K}} \sum_{ {x_i}\in ({{D_s}\cup{D_t}}) } \mathcal{L}^k_d(G^k_d(\hat{y}^k_i G_f(x_i; \theta_{f}); \theta_{d}), d_i)
    \end{aligned}
\end{equation}


where $n = n_s + n_t$ and $\lambda$ are coefficients that regulate the trade-offs among the adversarial objectives used to construct the domain-invariant feature during back-propagation. To determine the domain-invariant features and leverage the speaker-discriminative embeddings, our aim is to seek the best parameters $\theta_{f}$, $\theta_{y}$, and $\theta_{d}$ that minimize the label prediction loss while maximizing the domain prediction loss. After converging to a global optimum, the parameters $\hat{\theta}_f$, $\hat{\theta}_y$, $\hat{\theta^{k}_{d}}|^{K}_{k=1}$ will satisfy the following functional:

\begin{equation}
    \begin{aligned}
        (\hat{\theta}_f, \hat{\theta}_y)            & =
        \operatorname*{arg} \operatorname*{min}_{\theta_f, \theta_y} \mathcal{L} (\theta_f, \theta_y, \hat{\theta^{k}_{d}}|^{K}_{k=1}) \\
        (\hat{\theta}^1_d, {...}, \hat{\theta}^K_d) & =
        \operatorname*{arg} \operatorname*{max}_{\theta^1_d, {...}, \theta^K_d} \mathcal{L} (\hat{\theta}_f, \hat{\theta}_y, {\theta^{k}_{d}}|^{K}_{k=1})
    \end{aligned}
\end{equation}

Because the learned domain-invariant feature is domain specific, which is beneficial for its own domain and detrimental for other domains \cite{mao2020adaptive}, it is necessary to compare different domain discriminators and determine the optimal combination of feature extractor and discriminators.

\subsection{Integration of speaker-discriminative embeddings}

We investigated the feasibility of integrating multiple features for different speaker-discriminative embeddings. It should be possible to consider the different data distributions from each feature and then use the complementary information from different representations \cite{li2021feature}. We integrate the information of two representations through feature concatenation as follows:

\begin{equation}
    \begin{aligned}
        f =          & f_1 + f_2        \\
        f \leftarrow & append(f_1, f_2)
    \end{aligned}
\end{equation}

As shown in Fig. \ref{fig:arch}, two types of feature extractors do not share the parameters and generate two different representations from the same speech segment. We evaluated various acoustic features and choose the following feature extractors: ECAPA-TDNN \cite{desplanques2020ecapa}, which uses the filterbank energy feature as input, and RawNet2 \cite{jung2020improved}, which uses the raw waveform directly as input. We also use the ECAPA-TDNN pretrained model\footnote{\url{https://huggingface.co/speechbrain/spkrec-ecapa-voxceleb}} to avoid the problems that arise when NaN is obtained for the loss in the model trained from scratch.

\subsection{Multi-task learning architecture}

In this section, we describe our proposed network in the multi-task learning (MTL) scenario for speaker recognition and age group classification. MTL can handle multiple tasks concurrently through a learned shared representation and adaptively transfer useful information between task-specific networks. In our study, combining multiple tasks into a single architecture has an effect on the speed of model convergence to the global minimum. In addition, we show that using the shared representation with MTL improves the classification performance increasing the effectiveness of the adaptive adversarial network. As a result, the proposed method is capable of boosting the performance of multi-task learning models and is effective at aligning the distributions at domain level.

The basic expression for the MTL loss functions is given in Eq. \ref{eq:mtl_basic}, where $\mathcal{L}$ is the loss of the task $\tau$ and compares the groud-truth labels $y_{\tau}$ to predictions $y'_{\tau}$ to optimize learnable parameters ${\theta}_{\tau}$. Hyperparameter ${\mathcal{C}}_{\tau}$ is used to account for the differing variances and offsets in the single-task losses. We use automatic weighted loss (Eq. \ref{eq:mtl_awl}) to enable the model to automatically learn a weighting for the tasks that improves performance \cite{liebel2018auxiliary}. The model learns various quantities at different scales for the multiple classification tasks: a weight parameter ${\mathcal{C}}^2_{\tau}$ is used during back-propagation to enforce positive regularization values.

\begin{equation}
    \begin{aligned}
        \mathcal{L}_{MTL} (x, y_{\tau}, y'_{\tau}; {\theta}_{\tau}) =
        \sum_{ {\tau}\in{\mathcal{T}} } \mathcal{L}_{\tau} (x, y_{\tau}, y'_{\tau}; {\theta}_{\tau}) \cdot {\mathcal{C}}_{\tau}
    \end{aligned}
    \label{eq:mtl_basic}
\end{equation}

\begin{equation}
    \begin{aligned}
        \mathcal{L}_{\tau} (x, y_{\tau}, y'_{\tau}; {\theta}_{\tau}) =
        \sum_{ {\tau}\in{\mathcal{T}} } \frac{1}{2 \cdot \mathcal{C}^2_{\tau} }  \mathcal{L}_{\tau} (x, y_{\tau}, y'_{\tau}; {\theta}_{\tau}) + \ln(1+ {\mathcal{C}}^2_{\tau})
    \end{aligned}
    \label{eq:mtl_awl}
\end{equation}

As shown in Fig. \ref{fig:arch}, the AM softmax loss is computed for speaker recognition, and the age group classifier is trained using the cross-entropy loss. The loss used in the domain discriminators is the NLL loss with label smoothing for regularization \cite{pereyra2017regularizing}. The overall objective function is a weighted sum of all loss functions, where $\alpha$ is set to 0.01.

\begin{equation}
    \begin{aligned}
        \mathcal{L}_{total}  = \mathcal{L}_{\tau1}(x, y_{\tau1}, y'_{\tau1}; {\theta}_{\tau1}) + \alpha(\mathcal{L}_{\tau2}(x, y_{\tau2}, y'_{\tau2}; {\theta}_{\tau2}) \\
        \text{\scriptsize $\tau1 \in \{{\tau}_{spk}, {{\tau}_{ag}}\}, \quad  \tau2 \in \{{\tau}_{young}, {{\tau}_{adult}}, {{\tau}_{senior}} \}$}
    \end{aligned}
\end{equation}

\section{EXPERIMENTS}
\label{sec:EXPERIMENTAL SETUP}
\subsection{Dataset}

\begin{table}[!b]
    \centering
    \caption{Numbers of utterances and speakers in \textit{VoxCeleb Enrichment}: Y/A/S in order are young people, adult, senior}
    \resizebox{0.95\columnwidth}{!}{
        \begin{tabular}{c||ccc|ccc}
            \hline
            \multicolumn{1}{c||}{\multirow{2}{*}{\textbf{Datasets}}} & \multicolumn{3}{c|}{\textbf{leq 29 $^{\mathrm{a}}$}} & \multicolumn{3}{c}{\textbf{leq 17 $^{\mathrm{b}}$}}                                                                                  \\ \cline{2-4} \cline{5-7}
                                                                     & \textbf{\textit{\#train}}                            & \multicolumn{2}{c|}{\textbf{\textit{\#dev \#eval}}} & \textbf{\textit{\#train}} & \multicolumn{2}{c}{\textbf{\textit{\#dev \#eval}}} \\ \hline \hline
            \makecell{ \# utts                                                                                                                                                                                                                                     \\ (Y/A/S) }& \makecell{80,476 \\ 324,733 \\ 8,9235}                          &    \multicolumn{2}{c|}{\makecell{9,957 \\40,618 \\ 11,231}}                            &   \makecell{2,988 \\ 402,176 \\ 89,280}       &   \multicolumn{2}{c}{\makecell{355 \\ 50,181 \\ 11,270}}  \\ \hline
            \makecell{ \# spks                                                                                                                                                                                                                                     \\ (Y/A/S) }& \makecell{637 \\ 1,783 \\ 390}                          &    \multicolumn{2}{c|}{\makecell{635 \\ 1,778 \\ 390}}      &   \makecell{16 \\ 2,404 \\ 390}       & \multicolumn{2}{c}{\makecell{16 \\ 2,395 \\ 390}}   \\ \hline
            \multicolumn{6}{l}
            {\footnotesize {$^{\mathrm{a}}$ Y(${\leq29}$), A(${30\texttildelow60}$), S(${\geq60}$) \quad $^{\mathrm{b}}$ Y(${\leq17}$), A(${18\texttildelow60}$), S(${\geq60}$)}}
        \end{tabular}
    }
    \vspace{-2mm}
    \label{tbl:data_stats}
\end{table}

To evaluate our proposed method, we used the \textit{VoxCeleb Enrichment} dataset\cite{hechmi2021voxceleb} for the source and target domain data to train the model, and we used \textit{VoxCeleb1-H} to evaluate the performance of the unknown speakers \cite{nagrani2020voxceleb}. Because \textit{VoxCeleb Enrichment} and \textit{VoxCeleb1-H} were extracted from YouTube videos, the audio clips were recorded in a variety of acoustic environments. For the age group classification, the audio files were divided into three age groups: young people, adults, and seniors. Two conditions were used to divide the groups: for the call center voice response system data, young people are those less than or equal to 29 years in age (leq 29) \cite{sanchez2022age}, and for the Motion Picture Association (MPA) film rating system data, young people are those less than or equal to 17 years in age (leq 17). The statistics of the number of utterances and speakers in the dataset are shown in Table \ref{tbl:data_stats}.

\begin{figure}[t!]
    \centering
    \begin{minipage}{.45\linewidth}
        \subfigure{\includegraphics[width=0.9\linewidth,height=2.1cm]{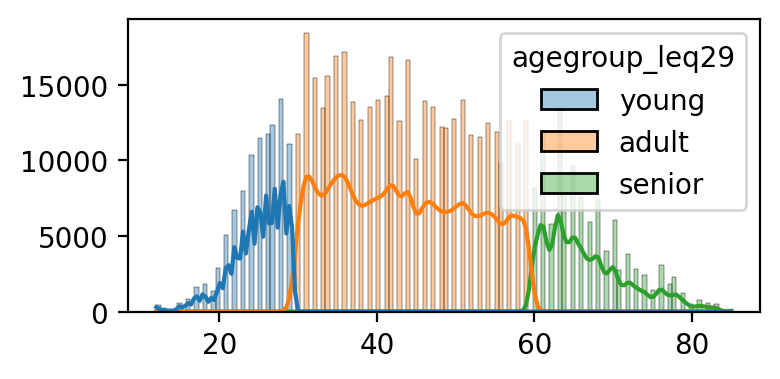}}
        \subfigure{\includegraphics[width=0.9\linewidth,height=2.1cm]{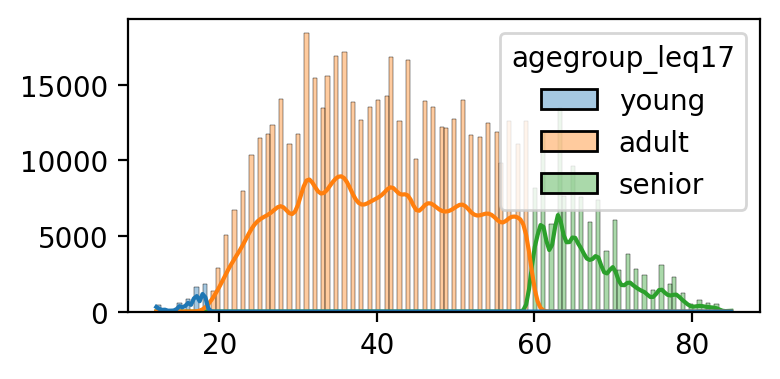}}
    \end{minipage}
    \begin{minipage}{.50\linewidth}
        \subfigure{\includegraphics[width=0.9\linewidth,height=4.5cm]{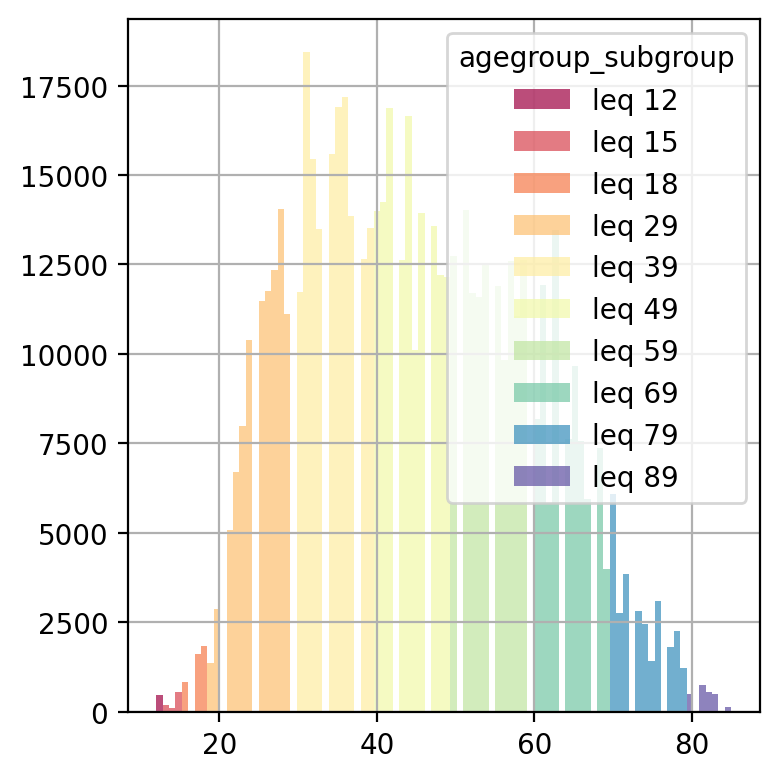}}
    \end{minipage}
    \caption{\textit{VoxCeleb Enrichment} dataset distribution for age group (left) and age subgroup (right)}
    \label{fig:dist_den}
\end{figure}

Fig. \ref{fig:dist_den} presents the distribution of the dataset used in our experiment. In this setting, the leq 29 division leads to relatively balanced label distribution, whereas the label distribution of the source domain for leq 17 is highly imbalanced. Note that because adaptive adversarial networks transfer the knowledge learned from a labeled source domain to an unlabeled target domain via statistical distribution alignment, the target-domain data are unlabeled. However, considering that the source and target domains originate from the features of the same speech, there are no unlabeled target-domain data in our work. Therefore, we grouped the age subgroups for each age group by referring to the age-by-decade and MPA film rating systems. Specifically, for an unlabeled target-domain sample, we divided the age group intervals as shown in Fig. \ref{fig:dist_den} (right).

\subsection{Experimental setup and training details}

We evaluated our approach under the following domain-discriminator conditions: without any discriminators (woD), with a young speaker discriminator (YD), with a senior speaker discriminator (SD), with a young and senior dual-discriminator (YSD), and with an adult discriminator (AD). All the experiments were conducted using PyTorch on two NVIDIA A100 GPUs. For preprocessing, each utterance was augmented by the speed perturbation methods of SpecAugment \cite{park2019specaugment} and converted to 80-dimensional Fbank features for E (ECAPA-TDNN) and raw waveform for R (RawNet2). After extraction, the dimensions of the speaker embeddings for the ECAPA-TDNN and RawNet were 256 and 512, respectively. The multi-task classifiers $G_y$ and the domain discriminators $G_d$ are both composed of FC layers with leaky ReLU activation, one-dimensional batch normalization, and dropout (0.5). To ensure effective training, we used an automatic mixed-precision technique, clipping gradient normalization with a maximum threshold of four, accumulated gradient for five batches, and a learning rate of 1e-4 that reduced by 20\% when the learning stagnated over two epochs.

\section{Results and Analysis}

Table \ref{tbl:results} summarizes the results of the evaluation set for age group classification. As expected, the MTL methods achieved better results than the single-task learning (STL) performance in age group classification. The results indicate that E-AD and E-R-YD produced the best average performance for age group classification with respect to the different domain discriminator conditions. However, the proposed adaptive adversarial network does not always outperform the woD condition. In addition, some of the combinations are not as effective as extracting the domain-invariant features for age group classification. The best age group prediction was achieved when optimizing $\tau \in \{{\tau}_{young}, {{\tau}_{adult}}, {{\tau}_{senior}}\}$, i.e., when choosing the appropriate age subgroup with the feature extractor. The performance results reveal that the target-domain data should be chosen carefully. Further experiments are needed to determine the age subgroup factor that supports the age group classification task needed to obtain better domain-invariant features.

\begin{table}[t!]\centering
    \caption{Performance of age group classification on the eval-set for various combinations of the feature extractors and the discriminators}
    \resizebox{7cm}{!}{
        \begin{tabular}{c|cc|ccc}
            \hline
            \multicolumn{1}{c|}{\multirow{2}{*}{\textbf{Dataset}}} & \multicolumn{2}{c|}{\multirow{2}{*}{\textbf{Methods}}} & \multicolumn{3}{c}{\textbf{Precision} [\%]}                                                                                \\
            \cline{4-6}
            \textbf{}                                              & \textbf{}                                              & \textbf{}                                   & \textbf{\textit{Young}} & \textbf{\textit{Adult}} & \textbf{\textit{Senior}} \\
            \hline \hline
            leq 29                                                 & STL                                                    & E-woD                                       & 44.89                   & 91.37                   & 53.88                    \\
            \hline
                                                                   &                                                        & E-woD                                       & 84.11                   & 97.93                   & 95.61                    \\
                                                                   &                                                        & E-YD                                        & 90.40                   & \textbf{99.62}          & 90.66                    \\
            leq 29                                                 & MTL                                                    & E-SD                                        & 88.77                   & 99.56                   & 94.51                    \\
                                                                   &                                                        & E-YSD                                       & 79.15                   & 98.60                   & 91.42                    \\
                                                                   &                                                        & \textbf{E-AD}                               & \textbf{94.76}          & 98.85                   & \textbf{97.31}           \\
            \hline
                                                                   &                                                        & E-R-woD                                     & 95.95                   & 99.34                   & 96.96                    \\
                                                                   &                                                        & \textbf{E-R-YD}                             & \textbf{99.29}          & \textbf{99.83}          & \textbf{99.65}           \\
            leq 29                                                 & MTL                                                    & E-R-SD                                      & 94.61                   & 98.23                   & 98.37                    \\
                                                                   &                                                        & E-R-YSD                                     & 93.55                   & 99.16                   & 94.58                    \\
                                                                   &                                                        & E-R-AD                                      & 95.28                   & 99.29                   & 96.85                    \\
            \hline \hline
            leq 17                                                 & STL                                                    & E-woD                                       & 24.39                   & 95.90                   & 65.36                    \\
            \hline
                                                                   &                                                        & E-woD                                       & 86.97                   & 99.79                   & 91.33                    \\
                                                                   &                                                        & E-YD                                        & 70.33                   & \textbf{99.82}          & 85.95                    \\
            leq 17                                                 & MTL                                                    & E-SD                                        & 79.17                   & 99.60                   & 93.42                    \\
                                                                   &                                                        & E-YSD                                       & 82.55                   & 99.71                   & 94.93                    \\
                                                                   &                                                        & \textbf{E-AD}                               & \textbf{87.95}          & 99.50                   & \textbf{96.00}           \\
            \hline
                                                                   &                                                        & E-R-woD                                     & 89.54                   & 99.87                   & 93.33                    \\
                                                                   &                                                        & \textbf{E-R-YD}                             & \textbf{93.58}          & 99.82                   & \textbf{97.03}           \\
            leq 17                                                 & MTL                                                    & E-R-SD                                      & 87.50                   & 99.88                   & 91.15                    \\
                                                                   &                                                        & E-R-YSD                                     & 87.94                   & \textbf{99.90}          & 92.67                    \\
                                                                   &                                                        & E-R-AD                                      & 81.59                   & 99.85                   & 96.04                    \\
            \hline
        \end{tabular}
    }
    \label{tbl:results}
\end{table}

\begin{figure}[b!]
    \centering
    \subfigure[(MTL) E-woD]{\includegraphics[width=0.32\linewidth]{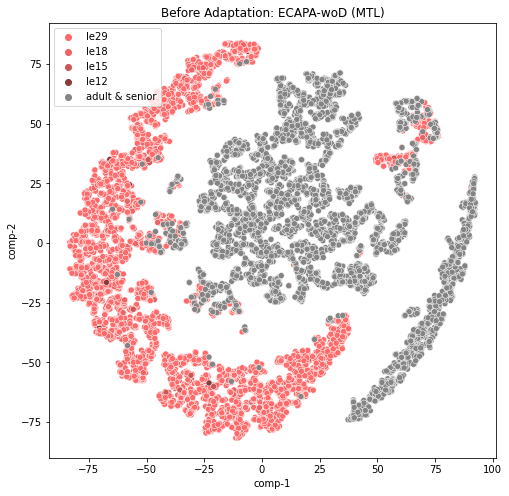}}
    \subfigure[(MTL) E-YD]{\includegraphics[width=0.32\linewidth]{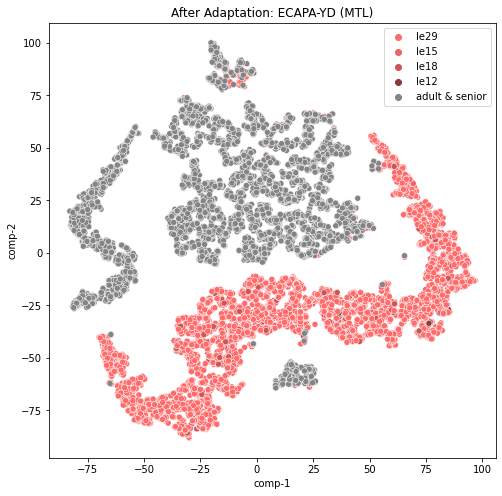}}
    \subfigure[(MTL) E-R-YD]{\includegraphics[width=0.33\linewidth]{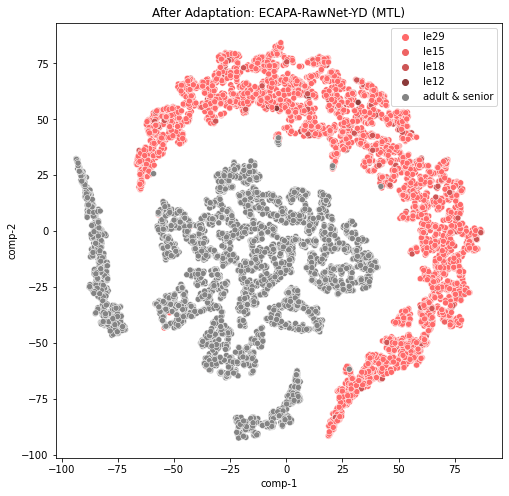}}
    \caption{
        t-SNE visualizations of the extracted embeddings on eval-set (leq 29): shades of red (young), grey (adult \& senior)
    }
    \label{fig:tsne_analysis}
\end{figure}

As shown in Table \ref{tbl:results}, the integration of the feature extractor (E-R) leads to high overall age group classification performance when compared with the results of the stand-alone feature extractor (E). The integrated speaker-discriminative embeddings provide better age group discriminability by enriching the information contained in the domain-invariant feature. However, when only considering the performance of speaker recognition on \textit{VoxCeleb1-H} (Table \ref{tbl:results_eer}), which is measured using metrics such as the equal-error rate (EER) and minimum detection cost function (min-DCF), this approach underperforms. In fact, the integration of embeddings is meaningless for identifying speakers, which are not specified in the source of the labeled training data. We determined that the relative strengths of the integrated features are that it enables the adaptation to concentrate on age group factors and encode known speakers, but it is difficult to analyze unknown speakers using this approach.

To visualize the effectiveness of the proposed method and illustrate the properties of the embeddings, t-distributed stochastic neighbor embedding (t-SNE) visualizations are presented in Fig. \ref{fig:tsne_analysis}. To observe the age group clusters for young people, half of the sample utterances were randomly chosen from the young speaker label, whereas the others were randomly chosen from the adult and senior speaker labels. It can be observed that the age group clusters for the young speakers in Fig. \ref{fig:tsne_analysis} (c) are more compact, whereas in Figs. \ref{fig:tsne_analysis} (a) and (b), there are small separated clusters for each age group.

\section{CONCLUSIONs}
\label{sec:CONCLUSION}

\begin{table}[t!]\centering
    \caption{Performance of speaker recognition on \textit{VoxCeleb1-H} and the eval-set for various combinations of feature extractors and discriminators. Cosine similarity is used for the scoring of positive and negative trials. 1.0 million utterances randomly sampled from eval-set are used for the trial pairs to evaluate EER and min-DCF.}
    \resizebox{0.95\columnwidth}{!}{
        \begin{tabular}{c|cc|cc|cc}
            \hline
            \multicolumn{1}{c|}{\multirow{2}{*}{\textbf{Dataset}}} & \multicolumn{2}{c|}{\multirow{2}{*}{\textbf{Methods}}} & \multicolumn{2}{c|}{\textbf{VoxCeleb1-H}} & \multicolumn{2}{c}{\textbf{Eval}}                                                                                    \\
            \cline{4-7}
            \textbf{}                                              & \textbf{}                                              & \textbf{}                                 & \textbf{\textit{EER}} [\%]        & \textbf{\textit{minDCF}} & \textbf{\textit{EER}} [\%] & \textbf{\textit{minDCF}} \\
            \hline \hline
                                                                   &                                                        & E-woD                                     & 2.57                              & 0.168                    & 0.94                       & 0.031                    \\
                                                                   &                                                        & E-YD                                      & 2.33                              & 0.172                    & 1.04                       & 0.035                    \\
            leq 29                                                 & MTL                                                    & E-SD                                      & 2.59                              & 0.176                    & 1.07                       & 0.042                    \\
                                                                   &                                                        & E-YSD                                     & 2.64                              & 0.202                    & 1.01                       & 0.038                    \\
                                                                   &                                                        & \textbf{E-AD}                             & \textbf{2.31}                     & \textbf{0.148}           & \textbf{0.22}              & \textbf{0.012}           \\
            \hline
                                                                   &                                                        & E-R-woD                                   & 27.28                             & 1.000                    & 3.75                       & 0.141                    \\
                                                                   &                                                        & \textbf{E-R-YD}                           & \textbf{3.29}                     & \textbf{0.250}           & \textbf{0.82}              & \textbf{0.030}           \\
            leq 29                                                 & MTL                                                    & E-R-SD                                    & 6.68                              & 0.371                    & 1.03                       & 0.034                    \\
                                                                   &                                                        & E-R-YSD                                   & 26.09                             & 1.000                    & 2.76                       & 0.068                    \\
                                                                   &                                                        & E-R-AD                                    & 23.19                             & 1.000                    & 1.59                       & 0.056                    \\
            \hline \hline
                                                                   &                                                        & E-woD                                     & 2.77                              & 0.188                    & 0.84                       & 0.031                    \\
                                                                   &                                                        & E-YD                                      & 2.34                              & 0.169                    & \textbf{0.74}              & 0.027                    \\
            leq 17                                                 & MTL                                                    & E-SD                                      & 2.57                              & 0.188                    & 0.83                       & 0.030                    \\
                                                                   &                                                        & E-YSD                                     & 2.69                              & 0.194                    & 0.79                       & 0.029                    \\
                                                                   &                                                        & \textbf{E-AD}                             & \textbf{1.93}                     & \textbf{0.153}           & 0.85                       & \textbf{0.026}           \\
            \hline
                                                                   &                                                        & E-R-woD                                   & 17.06                             & 1.000                    & 2.19                       & 0.078                    \\
                                                                   &                                                        & \textbf{E-R-YD}                           & \textbf{5.26}                     & \textbf{0.288}           & \textbf{1.06}              & \textbf{0.039}           \\
            leq 17                                                 & MTL                                                    & E-R-SD                                    & 19.83                             & 1.000                    & 2.90                       & 0.119                    \\
                                                                   &                                                        & E-R-YSD                                   & 20.09                             & 0.999                    & 1.61                       & 0.048                    \\
                                                                   &                                                        & E-R-AD                                    & 28.72                             & 1.000                    & 2.76                       & 0.070                    \\
            \hline
        \end{tabular}
    }
    \label{tbl:results_eer}
\end{table}

To improve the performance of speaker age group classification, we presented a new concept for increasing the suitability of speaker-discriminative embeddings by adapting age subgroups using an adaptive adversarial network. Experimental results demonstrate the ability of the proposed adversarial multi-task learning methods to leverage speaker embeddings and determine domain-invariant features for domain-specific tasks. In addition, embedding integration is advantageous for deeply analyzing the features of an age group. Although only two types of speaker embeddings were used for the integration in this study, the integration methods could be easily replaced with a different combination of feature extractors. In future work, it would be interesting to explore the limitations of adaptive adversarial networks and analyze the class-level alignment for the age group label itself. In addition, the two main ideas proposed in this study could be extended to other applications of speech information, such as language and emotion classification, by leveraging and applying speaker-discriminative embeddings.

\clearpage

\ninept
\bibliographystyle{IEEEbib}
\bibliography{refs}
\end{document}